\documentclass[11pt]{article}
\usepackage{amsfonts,amssymb,amsmath}

\usepackage[english]{babel}
\usepackage{color}

\newtheorem{theorem}{Theorem}
\newtheorem{definition}{Definition}

\newtheorem{proof}{Proof}

\newtheorem{proposition}{Proposition}
\newtheorem{lemma}{Lemma}

\newcommand{\beq}{\begin{eqnarray}}
\newcommand{\eeq}{\end{eqnarray}}

\newcommand{\beqt}{\begin{eqnarray*}}
\newcommand{\eeqt}{\end{eqnarray*}}

\newcommand{\be}{\begin{equation}}
\newcommand{\ee}{\end{equation}}

\newcommand{\bl}{\begin{lemma}}
\newcommand{\el}{\end{lemma}}

\newcommand{\bt}{\begin{theorem}}
\newcommand{\et}{\end{theorem}}

\newcommand{\bd}{\begin{definition}}
\newcommand{\ed}{\end{definition}}

\newcommand{\bp}{\begin{proposition}}
\newcommand{\ep}{\end{proposition}}

\newcommand{\bpr}{\begin{proof}}
\newcommand{\epr}{\end{proof}}

\newcommand{\bi}{\begin{itemize}}
\newcommand{\ei}{\end{itemize}}

\newcommand{\ben}{\begin{enumerate}}
\newcommand{\een}{\end{enumerate}}

\begin{document}

\title{{\bf The roles of random boundary conditions  in spin systems}}
 
\author{ Eric O. Endo \footnote{NYU-ECNU, Institute of Mathematical Sciences at NYU  Shanghai, Shanghai,  China,
 \newline 
email: eoe2@nyu.edu}\\
Aernout C.D.  van Enter \footnote{ Bernoulli Institute, University of Groningen, Nijenborgh 9, 9747AG, Groningen, Netherlands,
 \newline
 email: aenter@phys.rug.nl}\\ 
  Arnaud Le Ny \footnote{LAMA UMR CNRS 8050, UPEC, Universit\'e Paris-Est,  94010 Cr\'eteil, France,
 \newline
 email:  arnaud.le-ny@u-pec.fr}
}

\maketitle

\begin{center}
{\bf Abstract:} 
\end{center}

Random boundary conditions are one of the simplest realizations of quenched disorder. They have been used as an illustration of various conceptual issues in the theory of disordered spin systems. Here we review some  of these results.

\footnotesize

\vspace{3cm}



\normalsize
\section{Introduction}

 In the theory of disordered systems, and in particular in the theory of spin glasses, for which the existence of phase transitions is so far unproven and the nature of the conjectured transition even among theoretical physicists is a topic of controversy, the traditional approach of selecting different Gibbs states by imposing boundary conditions geared towards a preferred state is ineffective, as we don't know which (Gibbs or ground) preferred states there could be to select from.
 
  Physically, it then makes more sense considering boundary conditions which are independent of the interactions. In the case of spin glasses those could be fixed, periodic or free, for example. Indeed, choosing boundary conditions which depend on a realization of a set of random interactions is not a physically feasible procedure. For an early discussion of this point we refer to \cite{EF}. 
 
 The mathematical theory of disordered spin systems has been described for example in \cite{Bov,New, Pet}. See also \cite{NS1} for a reader-friendly introduction to the issues which show up in the spin-glass problem.

 Having  boundary conditions independent of the interactions has played a role in the proper definition of spin-glass Edwards-Anderson order parameters \cite{CE,EG} and also in the issue of weak versus strong uniqueness of Gibbs measures \cite{CCST,CE,GNS}. 
 
 A possibility which can naturally occur in disordered models is non-convergence of finite-volume Gibbs measures in the thermodynamic limit (``Chaotic Size Dependence'') \cite{Esg,NS2}. If the finite-volume states don't converge, it still might be the case that distributional limits exist. Such limiting objects then are called ``metastates''. Tractable examples of them, one has to either consider mean-field models, see e.g. \cite{BG,IK}, but for short-range lattice models one usually needs to consider  somewhat simplified models, as e.g. in \cite{ENS1,ENS2}. 
 
 Considering deterministic models with random boundary conditions can provide suitable illustrations of various conceptual issues. For some  descriptions of the analogy between spin glasses with fixed boundary conditions  and deterministic models with random boundary conditions, see \cite{BCC, CCST, Esg,NS2}. Short-range Ising models have been studied in \cite{ENS1, ENS2}, and more recently one-dimensional long-range models have been considered \cite{EEL}. Here we review some of the results which were found in those examples.  

\section{Background and Notation on Disordered Spin Models}
\subsection {Spin Models and  Disorder}
 
 We will consider spin models in which we denote  spin configurations (respectively spins at site $i$)  by $\sigma$ (respectively $\sigma_i$), on a connected, infinite and locally-finite graph $G$. The state space is $\Omega_0$, the spin configuration space is $\Omega^G_0$. 

For a subset of vertices $G' \subset G$, we denote by $\sigma_{G'}=(\sigma_i)_{i\in G'}$ the configuration restricted on $G'$. Define $\Omega_0^{G'}$ to be the set of configurations on $G'$.
 
Let $\Phi=(\Phi_X)_{X\subset G,\ X \text{ finite}}$ be a family of $\mathcal{F}_X$-measurable functions $\Phi_X:\Omega_0^G\to \mathbb{R}$, where $\mathcal{F}_X$ is the local $\sigma$-algebra generated by the cylinders $\sigma_X$;   we will call such a family an  interaction.

 Given a finite-volume $\Lambda \subset G$, finite-volume Hamiltonians are expressed in terms of interactions
 $$
 H_{\Lambda}(\sigma_{\Lambda} b_{\Lambda^c}) = \sum_{X \cap \Lambda \neq \emptyset} \Phi_X (\sigma_{\Lambda} b_{\Lambda^c}).
 $$
 Here $b$ is an arbitrary boundary condition, an element of $\Omega_0^{G}$, identified with its projections 
 $\Omega^{\Lambda^c}_0$, and
 $$
 \sigma_{\Lambda} b_{\Lambda^c}=
 \begin{cases}
 \sigma_i, &\text{ if }i\in \Lambda,\\
 b_i, & \text{ if }i\in \Lambda^c.
 \end{cases}
 $$
  
 From such Hamiltonians one constructs finite-volume Gibbs measures on volume $\Lambda$,  with boundary condition $b$ on $\Lambda^c$ and inverse temperature $\beta>0$,
 $$
\mu_{\Lambda,\beta}^{b}(\sigma_{\Lambda}) = 
\frac{1}{Z_{\Lambda,\beta}^{b}} e^{-\beta H_{\Lambda}(\sigma_{\Lambda} b_{\Lambda^c})}.
$$
The normalization
$$
Z_{\Lambda,\beta}^{b} = \sum_{\sigma\in \Omega_0^{\Lambda}}e^{-\beta H_{\Lambda}(\sigma_{\Lambda} b_{\Lambda^c})}
$$
is called partition function. Under appropriate summability conditions on the interaction, in the thermodynamic limit (infinite-volume) Gibbs measures exist, also known as DLR measures. For the theory of infinite-volume Gibbs measures we refer to \cite{EFS,FV,HOG, Rue}.

 Although the theory applies in wider generality, we will tend to restrict ourselves to Ising spins $\sigma_i \in \Omega_0=\{-1,1\}$.

 Disordered systems depend on another random  parameter, the disorder parameter $\eta$. This disorder parameter can describe either bond randomness  or site randomness in the interactions, which then become random $\mathcal{F}_X$-measurable functions  $\Phi_X^{\eta}(\cdot)$ for each $X$ finite. Usually the variables $\eta$ are independent random variables with a distribution which is translation invariant and which depends on the shapes of the subsets of the lattice $X$. 
 
 There exists an extensive literature, both in (rigorous and nonrigorous) theoretical  physics and mathematical physics  of disordered systems. Here we refer to \cite{Bov,New, NS1, Pet} for some further mathematical and conceptual background and theory on them.
 
 We warn the reader, moreover, that the well-known random-bond equivalent-neighbour Sherrington-Kirkpatrick model of a spin-glass, although it has been rigorously solved by Guerra and Talagrand, following the ideas of Parisi, in many aspects is  exceptional and many statements which apply to it have no equivalent statement in the context we discuss. For some of the arguments on these issues, see \cite{Bov,New,NS1,NS2,NS3,NS4, Tal1,Tal2}. 

\subsection{ Examples. From spin glass to Mattis disorder to random boundary conditions.} 
Popular examples of disordered Ising systems include:

\begin{enumerate}
\item[1)]  {\bf Edwards-Anderson spin-glasses:} on $G=\mathbb{Z}^d$ with Hamiltonian
$$
H_{\Lambda}(\sigma_{\Lambda}b_{\Lambda^c})= - \sum_{\substack{i,j\in \Lambda \\ i\neq j}} \eta_{i,j} J(i-j) \sigma_i \sigma_j- \sum_{\substack{i\in \Lambda \\ j\in \Lambda^c}} \eta_{i,j} J(i-j) \sigma_i b_j,
$$
where the distribution of the (bond-random) $\eta_{i,j}$ is symmetric and depends only on $|i-j|$. They are usually taken as centered Gaussian or symmetric $\pm 1$. 
\item[2)] {\bf Random-field Ising models:} on $G=\mathbb{Z}^d$ with Hamiltonian
$$
H_{\Lambda}(\sigma_{\Lambda}b_{\Lambda^c})= - \sum_{\substack{i,j\in \Lambda \\ i\neq j}} J(i-j) \sigma_i \sigma_j - \sum_{\substack{i\in \Lambda \\ j\in \Lambda^c}}J(i-j) \sigma_i b_j - \lambda \sum_{i\in \Lambda} \eta_i \sigma_i,
$$
where the $\eta_i$ are (site-random) i.i.d. and symmetrically distributed random variables. Just as with Edwards-Anderson models, the most considered distributions are centered Gaussian and Bernoulli distributions. 
\item[3)] {\bf Mattis spin glasses:} on $G=\mathbb{Z}^d$ with Hamiltonian
$$
H_{\Lambda}(\sigma_{\Lambda}b_{\Lambda^c}) = - \sum_{\substack{i,j\in \Lambda \\ i\neq j}} J(i-j) \eta_i \eta_j \sigma_i \sigma_j - \sum_{\substack{i\in \Lambda \\ j\in \Lambda^c}} J(i-j) \eta_i \eta_j \sigma_i b_j,
$$
where again the (site-random) $\eta_i$ are i.i.d. and symmetrically distributed, typically  $\pm 1$.
\end{enumerate}

So far the theory of Edwards-Anderson spin glasses lacks examples in which it is clear that phase transitions occur. See e.g. \cite{NS5}.

The random-field Ising models have phase transitions in case of nearest-neighbor models in dimension at least 3, and also in dimension 1 if we consider the long-range interaction with sufficiently slow decay. These results all agree with heuristic predictions, based on some form of an Imry-Ma argument \cite{IM}. In such an argument one compares the (free-)energy cost of an excitation due to the spin interactions with the energy gain due to the magnetic field term.

 This argument  has been rigorized in a number of cases, sometimes requiring a serious mathematical analysis
\cite{AW,BEEKR,BrK,COP,Litt1}.

The Mattis spin-glasses on the other hand, by the random gauge transformation $\sigma'_i = \eta_i \sigma_i$,  are equivalent to  ferromagnetic models. 
Thus the existence or not of a phase transition typically only requires  understanding the ferromagnet.  
We note that for a finite-volume Gibbs measure a fixed boundary condition by this random gauge transformation  is mapped to a random boundary condition \cite{NS2, NS3}.

We also notice that Mattis disorder is the same as single-pattern Hopfield disorder, which has been considered in particular in the mean-field version; for more on Hopfield models see e.g. \cite{Bov,BG}. 

\section{Earlier results and new heuristics on  random boundary conditions}

In the results which we review  below, we always impose boundary conditions which are drawn from a symmetric  i.i.d. product (Bernoulli) measure. This does not preselect the phase, and such boundary conditions are sometimes called ``incoherent'' (as introduced in \cite{NS4}, see also \cite{ENS2}). 

\subsection{Weak versus strong uniqueness}

We say that a model displays weak uniqueness of the Gibbs measure if for each choice of boundary condition almost surely (for almost all choices of the random interaction) the same infinite-volume Gibbs measure is approached.
Strong uniqueness holds if there exists a unique Gibbs measure for the model for almost all choices of the interaction.

It is known that one-dimensional high-temperature long-range spin-glass models display weak uniqueness without strong uniqueness \cite{FZ, GNS}.
  Other examples where this occurs are the nearest-neighbour Ising spin-glass models on a tree, between the critical temperature and the spin-glass temperature \cite{CCST}. 

Similarly to what happens in Mattis models, one can transform the disorder to the boundary, and in the temperature interval between the ferromagnetic transition $T_c$ (below which plus and minus boundaries produce different states in the thermodynamic limit) and the free-boundary-purity (or spin-glass) transition temperature $T_{SG}$, below which the limiting Gibbs measure obtained with free boundary conditions becomes non-extremal, there is weak uniqueness of the Gibbs measure without strong uniqueness.

Similar behaviour (weak but not strong uniqueness) has also been derived for a Potts-Mattis model on $\mathbb{Z}^d$ for high $q$, at the transition temperature, with $d \geq 2$ \cite{CE}.

\subsection{Nearest-neighbour Ising models at low temperatures, metastates}

We first summarize here the results derived and described  in  \cite{ENS1, ENS2} for the nearest-neighbour Ising model on $\mathbb{Z}^d$. If we consider an Ising model on a box of size $N^d$ with random boundary conditions, the ground state energy and also the low-temperature free energy satisfy a weak version of the local central limit theorem.  One obtains estimates for the probability of the boundary term of a boundary of size $N$ to lie, not in finite intervals (as in the proper local limit theorem) or in intervals of size $\sqrt N$  ( as in the ordinary Central Limit Theorem), but in intervals of size $N^{\delta}$, with some $\delta$ between $0$ and $\frac{1}{2}$. This still suffices to show that the probability that the boundary free energy is close to zero goes to zero at a fast enough  rate. 

This can be used to show that the boundary (free) energy in a reasonably precise way scales like $N^{\frac{d-1}{2}}$. From this it follows in particular that the (free) energy difference between plus and minus phase diverges, with large enough probability, and thus randomly one of the two tends to be preferred. 

The distributional limit behaviour can be described in terms of ``metastates'', objects which were introduced by Aizenman-Wehr \cite{AW} and Newman-Stein \cite{New, NS3,NS4} via  different constructions, which then were shown to be equivalent, see also \cite{CJK}.

 A metastate is a measure on Gibbs measures. In its support either extremal or  non-extremal Gibbs measures, or both, can occur. If the support of a metastate contains more than one measure, it is called ``dispersed''.

 In case the distribution is $\eta$-dependent, the translation covariant metastate in fact becomes a measure on measures (distributions) on Gibbs measures. Translation covariance here means that shifting the $\eta$ induces a shift of the corresponding random Gibbs measures in the metastate. 
 
The weight of a Gibbs measure in a metastate indicates the probability of finding that particular Gibbs measure for a randomly chosen volume for a particular realization of the interaction,  or else, the probability of finding that Gibbs measure for a random realization of the interaction in a given large volume. If the Gibbs measures are random, this metastate necessarily is also a random object.

Although the notion of metastate has been developed for spin glasses, these have turned out to be so intractable that most examples which could be  handled are either mean-field models with site-random variables (see for example \cite{BG,IK}), or other heavily simplified models.

In \cite{ENS1,ENS2} it was proven, for example,  that the metastates obtained with random boundary conditions live on the (extremal) plus and minus measures of the nearest-neighbour Ising model. Whereas the simple case of ground states with weak (finite-energy) boundary conditions --- that is, the bonds inside the volume are infinite, but boundary bonds are finite-- is mathematically fairly straightforward, the low-temperature case required a careful analysis, making use of the technique of  cluster expansions, including estimates on boundary contours, leading to a weak version of the local central limit theorem.  But the analysis ended up providing essentially the same result as holds at $T=0$.  As the weights in the metastate are obtained by exponentiating and then normalizing the boundary (free) energies, divergence of those boundary terms leads to weights which are either zero or one.

 When the sequence of volumes is sufficiently sparse, the plus and minus measures are in fact the only two limit points.  To prove this, one has to exclude null-recurrent behaviour when taking sequences of increasing volumes. By taking sparse sequences, this allows one to apply a Borel-Cantelli argument.

\subsection{Long-range Ising models, metastates}

In \cite{EEL} we have started to extend the analysis of the metastate description to one-dimensional long-range Ising systems. There has been a substantial progress in the study of such low-dimensional long-range Ising models, which are known to display phase transitions \cite{ACCN, BK, Dys,FILS,FrSp, I,IN,Joh, Kac}. For a number of more recent works on these  models see \cite{BHS,BFV,BEEL,BEEKR,CFMP,CMPR,CMP,COP,CELR, EKRS, ELN, MSV}. 

As our canonical example, on $G=\mathbb{Z}$ and $\alpha\in (1,2]$, we consider the Dyson models with Hamiltonian
$$
H_{\Lambda}(\sigma_{\Lambda}b_{\Lambda^c}) = - \sum_{\substack{i,j\in \Lambda \\ i\neq j}} |i-j|^{- \alpha} \sigma_i \sigma_j -  \sum_{\substack{i\in \Lambda \\ j\in \Lambda^c}} |i-j|^{- \alpha} \sigma_i \eta_j,
$$
where the site-random boundary $\eta_j$ are i.i.d., symmetrically distributed random variables on $\{-1,1\}$.

If we impose weak (finite-energy) random boundary conditions, on an interval of size $N$ (so boundary bonds are finite, but bonds inside the volume are infinite)  the ground states, which are the plus and minus configurations, have a difference in energy which scales as  $N^{\frac{3}{2} - \alpha}$ when $\alpha < \frac{3}{2}$, and is almost surely  bounded otherwise. This implies that for $\alpha > \frac{3}{2}$ the metastate lives on mixed ground state  measures, whereas for $\alpha < \frac{3}{2}$, similarly to what occurs in higher-dimensional short-range models, the metastate has only the plus and minus states in its support. For positive temperatures analogous results for metastates on Gibbs measures are expected and partially proven (\cite{EEL} in progress).

To prove that null-recurrence of the set of mixed states does not occur, in case the decay is slow enough, we again need to consider sufficiently sparse sequences of increasing volumes. Next to being needed for a Borel-Cantelli argument, this also allows us to treat the boundary energies of different volumes as (approximately) independent.

To obtain an  almost sure statement, a local-limit-type argument, along the lines  of  the one discussed  in Appendix B  of  \cite {ENS1} could then be invoked.

Denote by $\sigma^+_j=+1$ for all $j\in \mathbb{Z}$.
The main object we study is the formal expression
$$
W^+ = \sum_{i < 0} \sum_{j \geq 0} |i-j|^{- \alpha} \sigma^+_{j}\eta_i= \sum_{i <0} W_i.
$$
This expression describes the interaction of a random boundary condition on the negative half-line with the plus ground state configuration on the right half-line.

As the $W_i$ are independent, the expectation of each $W_i$ is zero, and the variance is $\text{Var}(W_i)= O(|i|^{2- 2 \alpha})$. 

Alternatively, we can write $W= \sum_{j \geq 0} W'_j$, with the random variables $W'_j$ having zero expectation, being strongly correlated and satisfying $\mathbb{E}( (W'_j)^2) = O(|j|^{1- 2 \alpha})$, so $(\text{Var}(W'_j))^{1/2} = O( |j|^{\frac{1}{2} - \alpha})$.
Now instead of the sum of the variances, we need to consider a sum of the  --non-independent--  $W_j$ themselves.

Therefore it follows that, whether one considers either a plus interval of size $N$ with a random 
boundary, or, alternatively,  a random interval of size $N$ with a plus boundary, both scale like $N^{\frac{3}{2} - \alpha}$. 

We remark that   the sum of left and right boundary energy terms on both sides of a large enough interval,  again can be written as a sum  of similar form and for that reason satisfies the same scaling. This provides the scaling of the boundary energies mentioned above.

The boundary terms of a finite interval consist of a left and a right boundary term, when the interval is large those can be treated as more or less independent.

Let $\mu^+_{\beta}$ and $\mu^-_{\beta}$ be the thermodynamic limit of the plus b.c. Gibbs measures $\mu^{+}_{\Lambda,\beta}$ and minus b.c. $\mu^{-}_{\Lambda,\beta}$, respectively.

We notice that if two boundary energies are only differing by a finite amount, the limiting Gibbs  measures (or ground state measures in the zero-temperature framework) are absolutely continuous with respect to each other. This happens almost surely when $\alpha > \frac{3}{2}$. In that case, $W$ is a well-defined, almost surely finite random variable with some non-trivial distribution. The weight distribution $\lambda=\lambda(\mu) \in [0,1]$ on mixed Gibbs measures $\mu$ given by
$$
\mu=\lambda \mu^+_{\beta} + (1-\lambda)\mu^-_{\beta},
$$ 
obtained by exponentiating the boundary energies and normalizing them, also has a non-trivial distribution. 

This means in particular that the measures in the support of this distribution  are different mixtures of the plus and minus states.  Therefore the metastate lives on different mixtures, rather than on pure states.

So far we have derived the results described above for low enough temperatures when $\alpha > \frac{3}{2}$, and for $T=0$ with finite boundary terms when $\alpha < \frac{3}{2}$. Just as in the nearest-neighbour case, the extension to positive temperatures in the second case requires a sophisticated low-temperature (contour expansion) analysis, which is in progress.

We remark moreover that  it follows from \cite{CJK} that a metastate supported on pure states also exists; however, its construction will have to be different than just imposing independent random boundary conditions (possibly by making use of a maximizing procedure, or by considering highly correlated boundary conditions). For the ferromagnet this is immediate, for the Mattis version of our models less so.

\section{Conclusion, Final Remarks}

Random boundary conditions for ferromagnets play a similar role as fixed boundary for spin glasses. In the case of spin glasses with Mattis disorder, there is in fact direct map between those two cases.

 Random boundary conditions  can be used to illustrate the concepts of weak and strong uniqueness, as well as  describing various metastate scenarios. In particular, in one-dimensional Dyson models with a decay power between $\frac{3}{2}$ and $2$, they lead in a natural way to examples where the apparently new phenomenon of dispersed metastates living on mixed Gibbs measures appear.

\newpage

{\bf Acknowledgments:}   
We dedicate this contribution to the memory of  Vladas Sidoravicius. His enthusiasm as well as  his broad  interests, which included  both  long-range models and  the new phenomena occurring in random versions of a wide class of statistical mechanics models, have often been a stimulus and inspiration for all of us. We thank our collaborators and discussion partners on this and  related projects, Rodrigo Bissacot, Loren Coquille, Roberto Fern\'andez, Bruno Kimura, Christof K\"ulske, Pierre Picco, Wioletta Ruszel, Cristian Spitoni and Evgeny Verbitskiy.

A.C.D. v.E. thanks Roberto Fern\'andez and NYU Shanghai for the invitation for a stimulating research  visit, during which  we finished our paper. This research is partly funded by the B\'{e}zout Labex, funded by ANR, reference ANR-10-LABX-58.

\end{document}